\documentclass[a4paper,11pt]{article}

\usepackage{amsmath,amssymb}     
\usepackage{color}
\usepackage{graphicx}
\usepackage{subfigure}
\usepackage{cite}                
\usepackage{hyperref}            
\usepackage{multirow,makecell}   

\def \be {\begin{equation}}
\def \ee {\end{equation}}
\def \ba {\begin{array}}
\def \ea {\end{array}}
\def \bea{\begin{eqnarray}}
\def \eea{\end{eqnarray}}
\def \nn {\nonumber}

\def \m {\mu}

\def \O {\Omega}

\def \mF {\mathcal F}

\def \mR {\mathcal R}

\def \p {\partial}
\def \f {\frac}

\def \ma {\mathcal}
\def \mc {\mathcal}

\def \lt {\left}
\def \rt {\right}

\def \sr {\sqrt}

\begin{document}

\title{\textbf{Thermodynamics of two-dimensional conformal field theory dual to black holes}}
\author{
Jia-ju Zhang\footnote{jjzhang@pku.edu.cn}
}
\date{}

\maketitle

\vspace{-10mm}

\begin{center}
{\it
Department of Physics and State Key Laboratory of Nuclear Physics and Technology,\\Peking University, 5 Yiheyuan Road, Beijing 100871, China
}
\vspace{10mm}
\end{center}

\begin{abstract}

  In this note we investigate the first law of thermodynamics of the two-dimensional conformal field theory (CFT) that is dual to black holes. We start from the Cardy formula and get the CFT thermodynamics with minimal reasonable assumptions. We use both the microcanonical ensemble and canonical ensemble versions of the Cardy formula. In the black hole/CFT correspondence the black hole is dual to a CFT with excitations, and the black hole mass $M$ and charge $N$ correspond to the energy and charge of the excited CFT. The CFT left- and right-moving central charges $c_{L,R}$ should be quantized, and so we assume that they are mass-independent. Also we assume the difference of the left- and right-moving sector levels $N_L-N_R$ is mass-independent dual to level matching condition. The thermodynamics of two-dimensional CFT we get is universal and supports the thermodynamics method of black hole/CFT correspondence.

\end{abstract}

\baselineskip 18pt
\thispagestyle{empty}

\newpage

\tableofcontents

\section{Introduction}

How to understand the area law of the Bekenstein-Hawking entropy of the black hole \cite{Bekenstein:1973ur,Hawking:1974sw} in an microscopic way is one of the intriguing issues of quantum gravity.
The entropy for some kinds of extremal black holes in string theory could be reproduced by counting the degeneracy of brane configurations forming the black hole \cite{Strominger:1996sh}.
Also, One could use the fact that the quantum gravity in three-dimensional anti-de Sitter (AdS$_3$) spacetime is dual to a two-dimensional (2D) conformal field theory (CFT) \cite{Brown:1986nw}, and get the entropy of some black holes by counting the microstates of the near horizon geometry \cite{Strominger:1997eq}.
In recent years, this has been developed further to the study of ordinary rotating and/or charged black holes \cite{Guica:2008mu,Castro:2010fd}. Recent developments emphasize the role of the asymptotic symmetry group of the near horizon geometry of extremal black holes \cite{Guica:2008mu} and the hidden conformal symmetry in the scattering of a scalar off the nonextremal black holes \cite{Castro:2010fd}, instead of string techniques, to set up the correspondence.

It has been long observed that the mass-independence of the entropy product of the outer and inner horizons $S_+S_-$ and the inner mechanics might play important roles in the holographic descriptions of black holes \cite{Cvetic:1997uw,Cvetic:1997xv,Cvetic:2009jn,Cvetic:2010mn,Castro:2012av}, based on which there proposed the thermodynamics method of black hole/CFT correspondence proposed in \cite{Chen:2012mh,Chen:2012ps,Chen:2013rb}. It was found that many universal quantities of the dual CFT could be constructed using the thermodynamic quantities of the black hole outer and inner horizons. We will review the method below.
One of the essential assumptions in the method is the thermodynamics of the 2D CFT. In this note we will investigate the first law of thermodynamics of 2D CFT from the CFT itself. We start from the Cardy formula and get the CFT thermodynamics with minimal reasonable assumptions.
The thermodynamics of the CFT we get is universal and may apply to any 2D CFT with high excitations. This supports the thermodynamics method of black hole/CFT correspondence.

\section{CFT from black hole thermodynamics}

In the thermodynamics method of black hole/CFT correspondence we get the thermodynamic quantities of CFT from thermodynamics of the black hole outer and inner horizons \cite{Chen:2012mh,Chen:2012ps,Chen:2013rb}. Usually, the first law of the thermodynamics at the outer and inner horizons are
\bea \label{bhtherm}
&& d M=T_+ d S_+ + \O_+ d N,\nn\\
&& d M=-T_- d S_- + \O_- d N.
\eea
For simplicity we only have two quantities of the black hole to vary, i.e. the mass $M$ and one of the quantized charges $N$. Here $T_\pm$, $S_\pm$, $\O_\pm$ are the temperatures, entropies, and chemical potentials of the outer and inner horizons, respectively. The first laws characterize the responses of the black hole to perturbations. Thus the entropy product $S_+S_-$ being mass-independence is equivalent to the relation $T_+S_+=T_-S_-$. We may define
\be
\ma F \equiv \f{S_+ S_-}{4\pi^2}.
\ee
When $T_+S_+=T_-S_-$ is satisfied, $\mc F$ is a function of the charge $N$.

We recombine the black hole thermoeydnamics (\ref{bhtherm}) and get
\bea \label{CFTtherm1}
&& \f{1}{2}d M = T_L d S_L +\O_L d N,  \nn\\
&& \f{1}{2}d M = T_R d S_R +\O_R d N.
\eea
with the definitions \cite{Cvetic:1997uw,Cvetic:1997xv,Cvetic:2009jn,Chen:2012mh,Chen:2012ps,Chen:2013rb}
\bea
&& S_{L,R}=\f{1}{2}(S_+ \pm S_-), ~~~
   T_{L,R}=\f{T_-T_+}{T_- \mp T_+},   \nn\\
&& \O_{L}=\f{T_- \O_+ - T_+ \O_-}{2(T_- - T_+)}, ~~~
   \O_{R}=\f{T_- \O_+ + T_+ \O_-}{2(T_- + T_+)}.
\eea
We identify $T_{L,R}$, $S_{L,R}$ and $\O_{L,R}$ as temperatures, entropies and chemical potentials of the left- and right-moving sectors of the dual CFT, and thus (\ref{CFTtherm1}) as the first laws of thermodynamics of the CFT.

Furthermore, from (\ref{CFTtherm1}) we get
\be \label{CFTtherm2}
d N=T_L^N d S_L-T_R^N d S_R,
\ee
with
\be \label{R}
T_{L,R}^N=\mc R T_{L,R} , ~~~ \mc R=\f{1}{\O_R-\O_L}.
\ee
Here $\mc R$ has the dimension of length and is the characteristic length of the CFT. It is suggesting to identify $\mc R$ as radius of the circle the 2D CFT resides \cite{Cvetic:2009jn,Chen:2012mh}. The dimensionless quantities $T_{L,R}^N$ are left- and right-moving temperatures of the CFT in terms of the characteristic energy $\f{1}{\mc R}$.

Supposing the validity of the canonical ensemble version of the Cardy formula
\be
S_{L}=\f{\pi^2}{3}c_{L} T_{L}^N, ~~~
S_{R}=\f{\pi^2}{3}c_{R} T_{R}^N
\ee
we find that $T_+S_+=T_-S_-$ is equivalent to the equality of the left- and right-moving central charges
\be
c_L=c_R.
\ee
Moreover, in this case the central charges can be written as
\be \label{cLR}
c_{L,R}=6\f{\p \ma F}{\p N}.
\ee

For a black hole in diffeomorphism invariant gravity theory, its CFT dual is expected to be diffeomorphism invariant too. Thus we need $c_L=c_R$, and so $T_+S_+=T_-S_-$, or equivalently the mass-independence of the entropy product $S_+S_-$, is required for the black hole to have 2D CFT dual.
But for a black hole in gravity theory with diffeomorphism anomaly, its CFT dual has $c_L \neq c_R$, and so $S_+S_-$ needs to be mass-dependent. Examples are black holes in three-dimensional topologically massive gravity (TMG) \cite{Deser:1981wh,Deser:1982vy}, and the thermodynamics method still applies to these cases\cite{Chen:2013aza}. This is in accord with tracing the mass-dependence of $S_+S_-$ to the diffeomorphism anomaly of the theory \cite{Detournay:2012ug}.

\section{CFT thermodynamics}

In this section we investigate the thermodynamics in the CFT side and justify the assumptions in the last section. Our starting point is the microcanonical ensemble version of the Cardy formula \cite{Cardy:1986ie}
\be \label{cardy1}
S_{L}=2\pi \sr{\f{c_{L}N_{L}}{6}}, ~~~ S_{R}=2\pi \sr{\f{c_{R}N_{R}}{6}},
\ee
as well as the canonical ensemble version one
\be \label{cardy2}
S_{L}=\f{\pi^2}{3}c_{L} T_{L}^N, ~~~
S_{R}=\f{\pi^2}{3}c_{R} T_{R}^N.
\ee
Here $N_{L,R}$ are the levels of the excited state, and $T_{L,R}^N$ are the dimensionless temperatures. The validity of (\ref{cardy1}) needs $N_{L,R}$ to be large $N_{L,R} \gg c_{L,R}$, and the validity of (\ref{cardy2}) needs $T_{L,R}^N$ to be large $T_{L,R}^N \gg 1$. Supposing that there are regions of parameters where both (\ref{cardy1}) and (\ref{cardy2}) are valid, and then we can get \cite{Chow:2008dp}
\be
T_{L}^N=\f{1}{\pi}\sr{\f{6N_{L}}{c_{L}}}, ~~~ T_{R}^N=\f{1}{\pi}\sr{\f{6N_{R}}{c_{R}}}.
\ee
We discuss several cases.

\subsection{Case I}

The first case is the one-dimensional dilute gas in a circle with constant radius $\mc R$ \cite{Das:1996wn}. We suppose the left- and right-moving central charges are equal $c=c_{L,R}$, and $c$ is constant and does not depend on the energy $M$ or charge $N$. In this case there are simple relations
\be \label{e11}
N=N_L-N_R, ~~~ M=\f{N_L+N_R}{\mc R},
\ee
with $N_{L,R}$ being the quantized left- and right-moving momenta.
It can be checked easily that \cite{Chow:2008dp}
\be \label{e8}
d N_{L}=T_{L}^N d S_{L}, ~~~ d N_{R}=T_{R}^N d S_{R}.
\ee
Thus (\ref{CFTtherm2}) could be got, and also we could get (\ref{CFTtherm1}) with
\be
\O_{L,R}=\mp \f{1}{2 \mc R},
\ee
which is consistent to (\ref{R}). Now there is
\be
\mc F=\f{c N}{6},
\ee
and then (\ref{cLR}) is trivially satisfied.

Example is CFT dual to three-dimensional BTZ (Ba\~nados-Teitelboim-Zanelli) black hole \cite{Strominger:1997eq}. For CFT dual to BTZ black hole with mass $M$ and angular momentum $J$, there are the left- and right-moving entropies
\be
S_{L,R}=\pi \sr{\f{\ell(M\ell \pm J)}{2G}},
\ee
with $G$ being the Newton constant and $\ell$ being the AdS radius. So we have
\be
\mc F=\f{S_L^2-S_R^2}{4\pi^2}=\f{\ell}{4G}J,
\ee
which is mass-independent.
We choose $N=J$, and it is easy to see that the central charges and the radius of the circle the CFT resides are
\be
c_{L,R}=\f{3\ell}{2G}, ~~~ \mc R=\ell.
\ee
For CFT dual to five-dimensional Myers-Perry black hole with angular momenta $J_\phi$ and $J_\psi$, in the $J_\phi$ picture there is the central charges $c_{L,R}^\phi=6 J_\psi$ \cite{Lu:2008jk,Krishnan:2010pv}, which are $J_\phi$-independent, but there are no simple relations (\ref{e11}).

\subsection{Case II}

In the second case we suppose the central charge $c=c_{L,R}$ is mass-independent, but it may depend on the charge $N$. We define
\be
\mc F=\mc F_L -\mc F_R, ~~~ \mc F_{L,R}=\f{c N_{L,R}}{6}.
\ee
For level matching condition we suppose that $N_L-N_R$ is mass-independent,
\be \label{e9}
\f{\p N_L}{\p M} =\f{\p N_R}{\p M},
\ee
and so $\mc F$ is mass-independent too.
In this case we get
\bea \label{e10}
&& T_L^N dS_L=\f{\p N_L}{\p M} dM +\f{6}{c}\f{\p \mF_L}{\p N}dN,  \nn\\
&& T_R^N dS_R=\f{\p N_R}{\p M} dM +\f{6}{c}\f{\p \mF_R}{\p N}dN.
\eea
This is different to (\ref{e8}) dual to the $N$-dependence of $c$.
Using the relation (\ref{e9}), we get
\be
T_L^N dS_L-T_R^N dS_R=\f{6}{c}\f{\p \mF}{\p N} dN.
\ee
Since both $c$ and $\mF$ are functions of $N$ we can always get (\ref{CFTtherm2}) and(\ref{cLR}) by redefinition of $N$. We can also write (\ref{e10}) as
\bea
&& \f{1}{2}d M = \f{T_L^N}{\mR} d S_L +\O_L d N,  \nn\\
&& \f{1}{2}d M = \f{T_R^N}{\mR} d S_R +\O_R d N,
\eea
with the definitions
\bea
&& \mR=2\f{\p N_{L,R}}{\p M}, \nn\\
&& \O_{L,R}=-\f{6}{c \mR}\f{\p\mF_{L,R}}{\p N}.
\eea
It can be checked easily that
\be
\mR=\f{1}{\O_R-\O_L}.
\ee
These results are consistent with (\ref{CFTtherm1}) and (\ref{R}).

Examples are CFTs dual to four-dimensional Kerr black hole \cite{Guica:2008mu,Castro:2010fd}, four-dimensional Reissner-Nordstr\"om black hole \cite{Hartman:2008pb,Garousi:2009zx,Chen:2010as} and four-dimensional Kerr-Newman black hole \cite{Hartman:2008pb,Wang:2010qv,Chen:2010xu,Chen:2010ywa}. For CFT dual to four-dimensional Kerr black hole with mass $M$ and angular momentum $J$, there are
\be
S_L=2\pi GM^2, ~~~ S_R=2\pi \sr{G^2 M^4-J^2},
\ee
with $G$ being the Newton constant. So we have
\be
\mc F=\f{S_L^2-S_R^2}{4\pi^2}=J^2.
\ee
It is natural to choose the charge $N=J$ and get
\be
c_{L,R}=12J, ~~~ \mc R= \f{4G^2M^3}{J}.
\ee
However we can also choose $N=J^2$, which is still quantized, and then there would be
\be
c_{L,R}=6, ~~~ \mc R=8G^2M^3.
\ee

\subsection{Case III}

In the third case we suppose the central charges $c_L \neq c_R$, and both of them do not depend on the mass or charge. For level matching condition we suppose $N_L-N_R$ is mass-independent. In this case there are (\ref{e8}), and so we can define
\be
N=N_L-N_R
\ee
to get (\ref{CFTtherm2}). We can also get
\bea
&& \f{1}{2}d M = \f{T_L^N}{\mR} d S_L +\O_L d N,  \nn\\
&& \f{1}{2}d M = \f{T_R^N}{\mR} d S_R +\O_R d N,
\eea
with the definitions
\bea
&& \mR=2\f{\p N_{L,R}}{\p M}, \nn\\
&& \O_{L,R}=-\f{1}{\mR}\f{\p N_{L,R}}{\p N}.
\eea
It can be checked easily that
\be
\mR=\f{1}{\O_R-\O_L}.
\ee
These results are consistent with (\ref{CFTtherm1}) and (\ref{R}).

Since $c_L\neq c_R$, we can define
\be
c_{L,R}=c\pm \f{b}{2},
\ee
with $c$ being related to the conformal anomaly and $b$ being related to the diffeomorphism anomaly. We have
\be
\mF=\f{c_L N_L-c_R N_R}{6}=\f{c N}{6}+\f{b(N_L+N_R)}{12},
\ee
and so there are
\bea
&& b=\f{12}{\mR} \f{\p \mF}{\p M}, \nn\\
&& c=6  \f{\p\mF}{\p N} + \f{b \mc R}{2} (\O_L+\O_R) .
\eea

Examples are black holes in three-dimensional TMG, e.g. BTZ black hole and spacelike warped black hole \cite{Kraus:2005zm,Solodukhin:2005ah,Anninos:2008fx}.
For CFT dual to BTZ black hole in TMG with mass $M$ and angular momentum $J$, there are the left- and right-moving entropies
\be
S_{L}=\pi \sr{\lt( 1+\f{1}{\m\ell} \rt)\f{\ell(M\ell+J)}{2G}}, ~~~
S_{R}=\pi \sr{\lt( 1-\f{1}{\m\ell} \rt)\f{\ell(M\ell-J)}{2G}},
\ee
with $G$ being the Newton constant, $\ell$ being the AdS radius, and $\m$ being the TMG coupling constant. So we have \cite{Detournay:2012ug}
\be
\mc F=\f{S_L^2-S_R^2}{4\pi^2}=\f{\ell}{4G} \lt( J+\f{M}{\m} \rt),
\ee
which is mass-dependent.
We choose $N=J$, and it is easy to get
\be
c=\f{3\ell}{2G}, ~~~ b=\f{3}{G\m}, ~~~ \mc R=\ell.
\ee

\section{Conclusion}

In this note we investigated the thermodynamics of the 2D CFT that are dual to black holes in the CFT side. According to the known properties of CFTs dual to black holes we discussed three cases. The first case is a special one, the second one is more general, and the third case is for theory with diffeomorphism anomaly. Our derivation is quite general and we have only made minimal reasonable assumptions. There are three of basic assumptions in the note.
\begin{itemize}
  \item The validity of both the microcanonical ensemble (\ref{cardy1}) and canonical ensemble (\ref{cardy2}) versions of the Cardy formula.
  \item The central charges $c_{L,R}$ are mass-independent.
  \item For level matching condition, $N_L-N_R$ is mass-independent (\ref{e9}).
\end{itemize}
For all the cases we studied we can always get the thermodynamics of the CFT (\ref{CFTtherm1}) and (\ref{CFTtherm2}). This result supports the thermodynamics method of black hole/CFT correspondence. Moreover, the thermodynamics may be valid not only for CFTs dual to black holes, but also for general highly excited 2D CFTs.

\section*{Acknowledgments}

The author would like to thank Bin Chen for careful reading of the manuscript and valuable discussions and suggestions.
The work was in part supported by NSFC Grant No.~11275010, No.~11335012 and No.~11325522.

%



\providecommand{\href}[2]{#2}\begingroup\raggedright\endgroup

\end{document}